\documentclass[11pt]{article}
\usepackage{amssymb}

\begin{document}

\begin{center}
{\Large \bf Time Variation of the Fine Structure Constant \\ \vspace{0.2cm}
            in the Spacetime of a Cosmic Domain Wall}
\end{center}

\vspace{0.1cm}

\begin{center}
L. Campanelli$^{a,b,}$\footnote{E-mail: campanelli@fe.infn.it},
P. Cea$^{c,d,}$\footnote{E-mail: paolo.cea@ba.infn.it}, and
L. Tedesco$^{c,d,}$\footnote{E-mail: luigi.tedesco@ba.infn.it}
\end{center}

\begin{center}
$^{a}${\em INFN - Sezione di Ferrara, I-44100 Ferrara, Italy}, \\
$^{b}${\em Dipartimento di Fisica, Universit\`a di Ferrara, I-44100 Ferrara, Italy},
$^{c}${\em INFN - Sezione di Bari, I-70126 Bari, Italy}, \\
$^{d}${\em Dipartimento di Fisica, Universit\`a di Bari, I-70126 Bari, Italy}
\end{center}

\begin{center}
{ }
\end{center}

\vspace{1.0cm}

\begin{abstract}
The gravitational field produced by a domain wall acts as a medium
with spacetime-dependent permittivity $\varepsilon$. Therefore,
the fine structure constant $\alpha = e^2/4 \pi \varepsilon$ will
be a time-dependent function at fixed position. The most stringent
constraint on the time-variation of $\alpha$ comes from the
natural reactor Oklo and gives $|\dot{\alpha}/\alpha| < few \times
10^{-17} \, \mbox{yr}^{-1}$. This limit constrains the tension of
a cosmic domain wall to be less than $\sigma \lesssim 10^{-2} \,
\mbox{MeV}^{3}$, and then represents the most severe limit on the
energy density of a cosmic wall stretching our Universe.
\end{abstract}


\newpage

\section*{1. Introduction}

The physics of topological defects produced during cosmological
phase transitions has received a large amount of interest in
recent years. Topologically stable kinks are ensured when the
vacuum manifold of a spontaneously broken gauge theory is
disconnected \cite{Kibble76}. Let us consider, for simplicity, a
model in which kinks are infinitely static domain walls in the
$zy$-plane. That is we assume that the vacuum manifold consists of
just two disconnected components.

The dynamics and gravitational properties of such defects are
determined by their tension or surface energy density $\sigma$
\cite{{Vilenkin1},{vilenkin2}}. Unless the symmetry breaking scale
is very small, the surface density energy of the kink is extremely
large and implies that cosmic domain walls would have an enormous
impact on the homogeneity of the Universe. (Here and in the
following for cosmic domain walls we shall mean walls of linear
dimension $H_0^{-1}$, where $H_0$ is the Hubble constant). A
stringent constraint on the wall tension $\sigma$ for a cosmic
${\mathbb Z}_2\,$-wall can be derived from the isotropy of the
microwave background. If the interaction of walls with matter is
negligible, then there will be a few walls stretching across the
present horizon. They introduce a fluctuation in the temperature of
the microwave background of order $\delta T / T \simeq 2 \pi G
\sigma H_0^{-1}$ \cite{Zeld74}, where $G$ is the Newton's constant.
Observations constrain $\delta T / T \lesssim 3 \times 10^{-5}$, and
thus models predicting topologically stable cosmic walls with
$\sigma \gtrsim 1 \mbox{MeV}^3$ are ruled out.

In the following, we will see that the presence of a cosmic wall
stretching our Universe modifies the electromagnetic properties of
the free space. (This effect has been recently investigated in Ref.
\cite{{Nass1},{Nass2},{Nass3},{Nass4}} in the case of cosmic
strings.) In particular, the gravitational field produced by a wall
acts as a medium with time- and position-dependent permittivity.
This means that the fine structure constant $\alpha$, at fixed
position, will be a time-dependent function. Because terrestrial
experiments and observations constrain the time variation of
$\alpha$, we will be able to put a stringent limit on the energy
density of a cosmic wall.

\section*{2. The Fine Structure Constant in the Spacetime of a Domain Wall}

In this Section, we will see that the electric field generated by
a charge particle in the spacetime of a domain wall is the same as
in a flat spacetime but with a spacetime-dependent fine structure
constant.

We start by writing the line element associated to the spacetime
of a thin ${\mathbb Z}_2 \,$-wall \cite{Vil83}
\begin{equation}
\label{metric} ds^2 = e^{-4 \pi G \sigma |x|} (dt^2 - dx^2) - e^{4
\pi G \sigma (t - |x|)}(dy^2 + dz^2).
\end{equation}
Given a general diagonal metric
\begin{equation}
\label{eq1}
ds^2 = g_{\mu \nu} dx^{\mu} dx^{\nu} = g_{00} dt^2 - \gamma_{ij}
dx^i dx^j
\end{equation}
and the electromagnetic field strength tensor $F_{\mu \nu}$, the
electric and magnetic fields in a curved spacetime are defined as
\cite{Landau}
\begin{equation}
E_i = F_{0i}, \;\;\; B^i = -\frac{1}{2\sqrt{\gamma}} \,
\epsilon^{ijk} F_{jk},
\end{equation}
where $\gamma = \mbox{det} ||\gamma_{ij}||$ is the determinant of
the spatial metric and $\epsilon^{ijk}$ is the Levi-Civita symbol.
(Here and in the following, Greek indices run from $0$ to $3$, while
Latin indices run from $1$ to $3$.) The charge density of a particle
of charge $q$ at rest in the position ${\textbf x} = {\textbf x}_0$
is given by
\begin{equation}
\label{rho}
\rho = (q/\sqrt{\gamma}) \, \delta({\textbf x} - {\textbf x}_0).
\end{equation}
Introducing the fields
\begin{equation}
{\textbf D} = {\textbf E}/\sqrt{g_{00}}, \;\;\; {\textbf H} =
\sqrt{g_{00}}\, {\textbf B},
\end{equation}
the Maxwell's equations in three-dimensional notation read
\cite{Landau}
\begin{eqnarray}
\label{maxwell1}
&& \mbox{div} {\textbf B} = 0, \;\;\;\;\;\;\;
\mbox{curl} \, {\textbf E} = -\frac{1}{\sqrt{\gamma}}
\frac{\partial (\sqrt{\gamma} \,
{\textbf B})}{\partial t}, \\
\label{maxwell2}
&& \mbox{div} {\textbf D} = 4 \pi \rho, \;\;\;
\mbox{curl} \, {\textbf H} = \frac{1}{\sqrt{\gamma}}
\frac{\partial (\sqrt{\gamma} \, {\textbf D})}{\partial t},
\end{eqnarray}
where the divergence and curl differential operators are defined
in curved spacetime by
\begin{equation}
\label{xxx}
\mbox{div} \, {\textbf v} =
\partial_i (\sqrt{\gamma} v^i) / \sqrt{\gamma} \;
\end{equation}
and
\begin{equation}
\label{ccc}
\: (\mbox{curl} \, {\textbf v})^i = \epsilon^{ijk} (\partial_j v_k -
\partial_k v_j) / (2 \sqrt{\gamma}),
\end{equation}
respectively.
\\
It is convenient to re-write the first equation of
(\ref{maxwell2}) as
\begin{equation}
\label{Poisson} \nabla \cdot (\varepsilon {\textbf  E}) = 4 \pi q
\, \delta({\textbf x} - {\textbf x}_0),
\end{equation}
where $\nabla$ is the usual three-dimensional nabla operator in
Euclidean space, and we have introduced the parameter
$\varepsilon = \sqrt{\gamma}/\sqrt{g_{00}}$.
The solution of Poisson equation (\ref{Poisson}) is the standard
one:
\begin{equation}
\label{ttt}
\varepsilon {\textbf E} = (q/4 \pi r^3) \, {\textbf r},
\end{equation}
where ${\textbf r} = {\textbf x} - {\textbf x}_0$ and $r =
|{\textbf r}|$. Re-writing the above equation as
\begin{eqnarray}
\label{solution1} {\textbf E} = \frac{q}{4 \pi \varepsilon  r^3}
\, {\textbf r},
\end{eqnarray}
and taking into account the metric (\ref{metric}), we see that the
gravitational field produced by a domain wall acts as a medium
with permittivity $\varepsilon$ given by
\footnote{In the case of Taub metric \cite{Taub56} (i.e. the most
generic plane-symmetric metric) $ds^2 = e^{2u} (dt^2 - dx^2) -
e^{2v}(dy^2 + dz^2)$, where $u$ end $v$ are functions of $t$ and
$x$, the permittivity induced by the gravitational field is
$\varepsilon = e^{2v}$.}
\begin{equation}
\label{epsilon1} \varepsilon = e^{4 \pi G \sigma (t-|x|)}.
\end{equation}
In other words, the fine structure constant, defined in the free
space as $\alpha_0 = e^2/4 \pi$, becomes in the spacetime of a
domain wall
\begin{eqnarray}
\label{alpha} \alpha = \frac{e^2}{4 \pi \varepsilon} \, .
\end{eqnarray}

\section*{3. Discussion and Conclusions}

From the above analysis it results that, if a cosmic wall were
present within our Hubble horizon, then the fine structure
constant would be time- and position-dependent. In particular, at
fixed position, the time variation of $\alpha$ would be
\begin{equation}
\label{variation} \frac{\dot{\alpha}}{\alpha} = -4 \pi G \sigma.
\end{equation}
It is worthwhile noting that the ``effective'' variation of the
fine structure constant, Eq.~(\ref{variation}), is not in
contradiction with the Einstein Equivalence Principle which
implies that, locally in the spacetime, no variations of $\alpha$
can occur.
\footnote{It is well known \cite{Vilenkin} that the spacetime of a
domain wall is locally flat everywhere except at $x=0$. Therefore,
one can perform a coordinate transformation such that the line
element in Eq.~(\ref{metric}) becomes that of a flat spacetime
and, consequently, the Maxwell equations assume the ``classical''
form with $\varepsilon =1$. Then, in agreement with the Einstein
Equivalence Principle, no variation of the fine structure constant
occurs locally in any point of the spacetime of a domain wall
(excepting the points on the domain wall surface).}
Indeed, what is measurable in our case are only differences of
values of $\alpha$ calculated at different spacetime points. Say in
other words, only non-local variations of the fine structure
constant are physical. Concerning this, it should be noted that all
terrestrial experiments devoted to the detection of possible time
variations of $\alpha$ measure, indirectly, values of $\alpha$ at
different times. These terrestrial experiments set limits on the
time variation of $\alpha$ \cite{Uzan02}. Different experiments give
different constraints which, however, are in the narrow range
$|\dot{\alpha}/\alpha| < few \times 10^{-15} \, \mbox{yr}^{-1}$
\cite{{Marion},{Bize},{Fischer},{Peik}}. This, in turns, gives a
limit on the tension of a wall present in our Hubble volume, $\sigma
\lesssim 1 \mbox{MeV}^{3}$, which is of the same order of magnitude
of that resulting from the isotropy of the microwave background.
\\
The most stringent constraint on $\dot{\alpha}/\alpha$ comes from
the natural reactor Oklo \cite{Oklo} and is $|\dot{\alpha}/\alpha|
< few \times 10^{-17} \, \mbox{yr}^{-1}$ \cite{Petrov05}. This
limit constrains the tension of a cosmic wall to be less than
$\sigma \lesssim 10^{-2} \, \mbox{MeV}^{3}$, and then represents
the most severe limit on $\sigma$.

In conclusion, we have demonstrated that the gravitational field
produced by a domain wall acts as a medium with
spacetime-dependent permittivity and, consequently, the fine
structure constant $\alpha$ is a time-dependent function at fixed
position. Taking into account the most stringent constraint on the
time-variation of $\alpha$ coming from the natural reactor Oklo,
we derived an upper limit for the tension of a cosmic domain wall.
This represents the strongest upper limit on the energy density of
a cosmic wall stretching our Universe to date.

\vspace*{0.4cm}

L. C. thanks M. Giannotti for useful discussions.


\end{document}